\documentclass[journal=jacsat,manuscript=article]{achemso}
\usepackage[OT1]{fontenc}
\usepackage[usenames,dvipsnames]{color}
\usepackage[latin1]{inputenc}
\usepackage[english]{babel}
\usepackage{graphicx}
\usepackage{color}
\usepackage{amssymb,amsmath}
\usepackage[Gray,squaren]{SIunits}
\usepackage{xspace}
\usepackage{upgreek}
\usepackage{ulem}
\usepackage{epstopdf}
\usepackage[final]{pdfpages}

\author{Syed Abdullah Aljunid}
\author{Eng Aik Chan}
\author{Giorgio Adamo}
\author{Martial Ducloy}
\email{martial.ducloy@univ-paris13.fr}
\affiliation{Centre for Disruptive Photonic Technologies, TPI, Nanyang Technological University, 637371, Singapore.}
\alsoaffiliation{Laboratoire de Physique des Lasers, Universit\'{e} Paris 13, CNRS, (UMR 7538), F-93430, Villetaneuse, France.}
\alsoaffiliation{School of Physical and Mathematical Sciences, Nanyang Technological University, 637371 Singapore.}
\author{David Wilkowski}
\email{david.wilkowski@ntu.edu.sg}
\affiliation{Centre for Disruptive Photonic Technologies, TPI, Nanyang Technological University, 637371, Singapore.}
\alsoaffiliation{School of Physical and Mathematical Sciences, Nanyang Technological University, 637371 Singapore.}
\alsoaffiliation{Centre for Quantum Technologies, National University of Singapore, 117543 Singapore.}
\alsoaffiliation{MajuLab, CNRS-UNS-NUS-NTU International Joint Research Unit UMI 3654, Singapore.}
\author{Nikolay I. Zheludev}
\email{nzheludev@ntu.edu.sg}
\affiliation{Centre for Disruptive Photonic Technologies, TPI, Nanyang Technological University, 637371, Singapore.}
\alsoaffiliation{Optoelectronics Research Centre \& Centre for Photonic Metamaterials, University of Southampton, Southampton SO17}

\date{\today{}}
\title{Atomic Response in the Near-field of Nanostructured Plasmonic Metamaterial}
\abbreviations{}

\begin{document}
\begin{abstract}	
We report on reflection spectra of ceasium atoms in close vicinity of a nanostructured metallic meta-surface.  We show that the hyperfine sub-Doppler spectrum of the $6^2S_{1/2} - 6^2P_{3/2}$ resonance transition at $852$~nm is strongly affected by the coupling to the plasmonic resonance of the nanostructure. Fine tuning of dispersion and positions of the atomic lines in the near-field of plasmonic metamaterials could have uses and implications for the atom-based metrology, sensing and the development of atom-on-a-chip devices.
\end{abstract}

%

\maketitle

Since more than one century, physicists are analysing atomic spectra and confront their results with quantum theories. More recently, high precision atomic spectroscopic measurements and time keeping \cite{bloom2014optical,bize2005cold} are used for fundamental test in physics \cite{bize2005cold,tsigutkin2009observation,griffith2009improved}. However, characteristics of a transition are not an inherent property of the atom but are also influenced by the environment. In particular, the spontaneous emission rate and lineshapes depend on the presence of a nearby media, waveguides and resonators \cite{kleppner1981inhibited,poddubny2013hyperbolic,dintinger2005strong,harding2015nanophotonic,spillane2008observation,stern2013nanoscale,sague2007cold,stehle2014cooperative,stern2014fano,ballin2013three}. These environmental dependencies could be a limitation for precision measurements, but becomes advantages for applications such as quantum information processing \cite{haroche2006exploring}. Beyond atoms, hybridizations of more complex atomic-like systems such as quantum dots and carbon nano tubes with nanowires and plasmonic metamaterials have been reported \cite{akimov2007generation,chang2007single,tanaka2010multifold,nikolaenko2010carbon,zheludev2012metamaterials}. Hybridization of atoms with metamaterial leading to strong atom-cavity coupling, line shaping, spontaneous emission modification and lattice interactions were also investigated theoretically \cite{klimov2002spontaneous,gupta2010signature,chakrabarti2010controlling,chakrabarti2009switching}.

\begin{figure}
\begin{center}
\includegraphics[width = 0.3\textwidth]{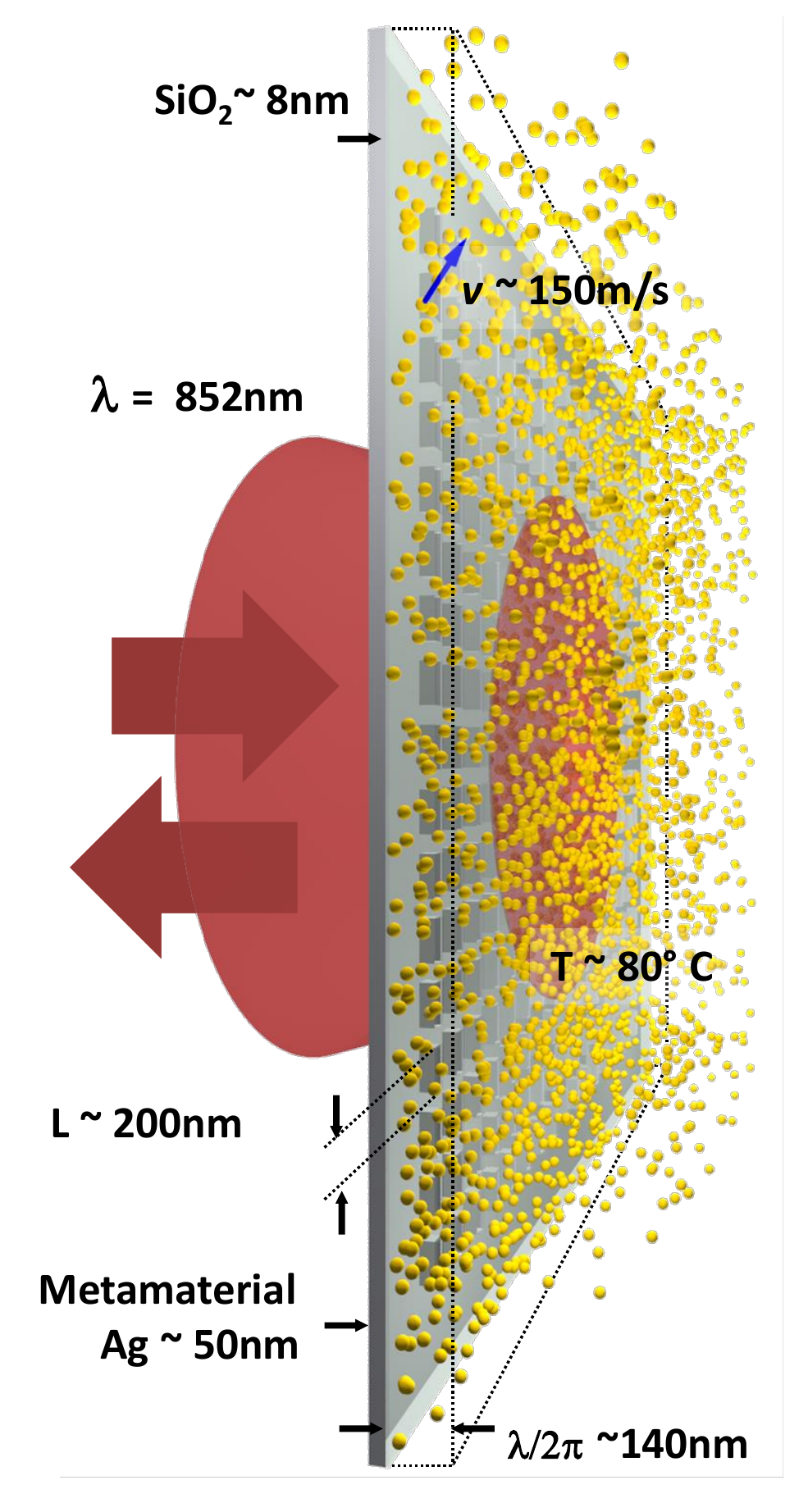}
\caption{Probing the sub-Doppler spectra of the caesium vapour at the vicinity of nanostructured plasmonic metamaterial. The yellow dots represent the atoms of the thermal vapour. Only the atoms propagating along the surface (blue arrow) within a layer of $\lambda/2\pi$ thickness are detected.}
\label{fig1}
\end{center}
\end{figure}

Here we report on a new type of hybrid system where atomic gas is hybridized with a plasmonic metamaterial (Fig. \ref{fig1}). In this extreme Fano system \cite{luk2010fano}, the plasmonic linewidth is seven orders of magnitude greater than the natural linewidth of atomic absorption lines. Nevertheless, in the regime of a few caesium atoms per unit cell located and probed in the near-field of the metamaterial, we observe strong plasmo-atomic coupling resulting in profound modifications of the shape of the atomic lines.

The hybrid metamaterial under study consists of an array of nano-slits in a silver film manufactured at the internal surface of optical window of a caesium vapour cell (Fig. \ref{fig1}). Our research focuses on the interaction between hyperfine transitions of the main dipole allowed D$_2$ line ($6^2S_{1/2} - 6^2P_{3/2}$,  $\lambda = 852$~nm) of caesium (Fig. 2) and dipolar plasmonic modes supported by the slit-type metamaterials. Here external reflectance measurements give access to the sub-Doppler spectrum of the atomic caesium transition that has natural lifetime of $30.5$~ns (linewidth  $\gamma_a = 2\pi\times 5.2$~MHz) \cite{nienhuis1988nonlinear}. Indeed the phase of the field back-reflected by the atoms depends on their distance from the interface and their contribution to the reflectivity rapidly average to zero for atoms located farther away from the interface. In practice, only a layer of gas of about $d = k^{-1} \simeq 136$~nm thickness, on top of the metamaterial, contributes significantly to the reflectivity. Here $k$ is the wavenumber of the probing optical wave. Then, atoms with substantial component of thermal velocity $\bar{v} \simeq 150$~m/s that is normal to the interface rapidly travel through the active layer with characteristic transient time of $1/k\bar{v} \sim 0.9$~ns that is much shorter than natural lifetime of the transition and therefore do not contribute to the sub-Doppler spectra \cite{akul1982collisional,ducloy1991general}. Moreover, atoms flying along the interface contribute to the reflectance spectra with no Doppler broadening as their velocities have no projection on the light wave-vector direction.

\begin{figure}
\begin{center}
\includegraphics[width = 0.35\textwidth]{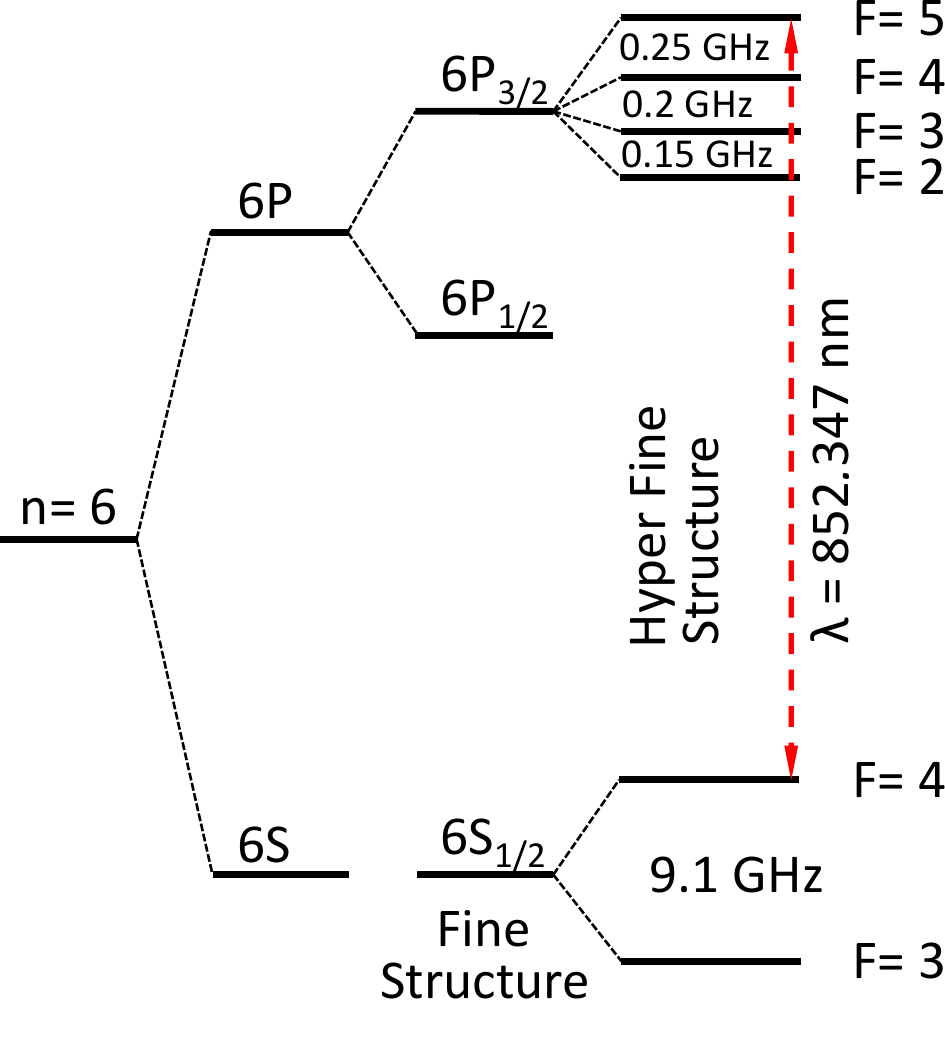}
\caption{Fine and hyperfine structure of the atomic caesium D$_2$ line. $6^2S_{1/2} \ F = 4 - 6^2P_{3/2} \ F=5$  hyperfine transition (in dotted red arrow line) is targeted for spectroscopy of the atom-metamaterial interaction.   }
\label{fig2a}
\end{center}
\end{figure}

\begin{figure}
\begin{center}
\includegraphics[width = 1\textwidth]{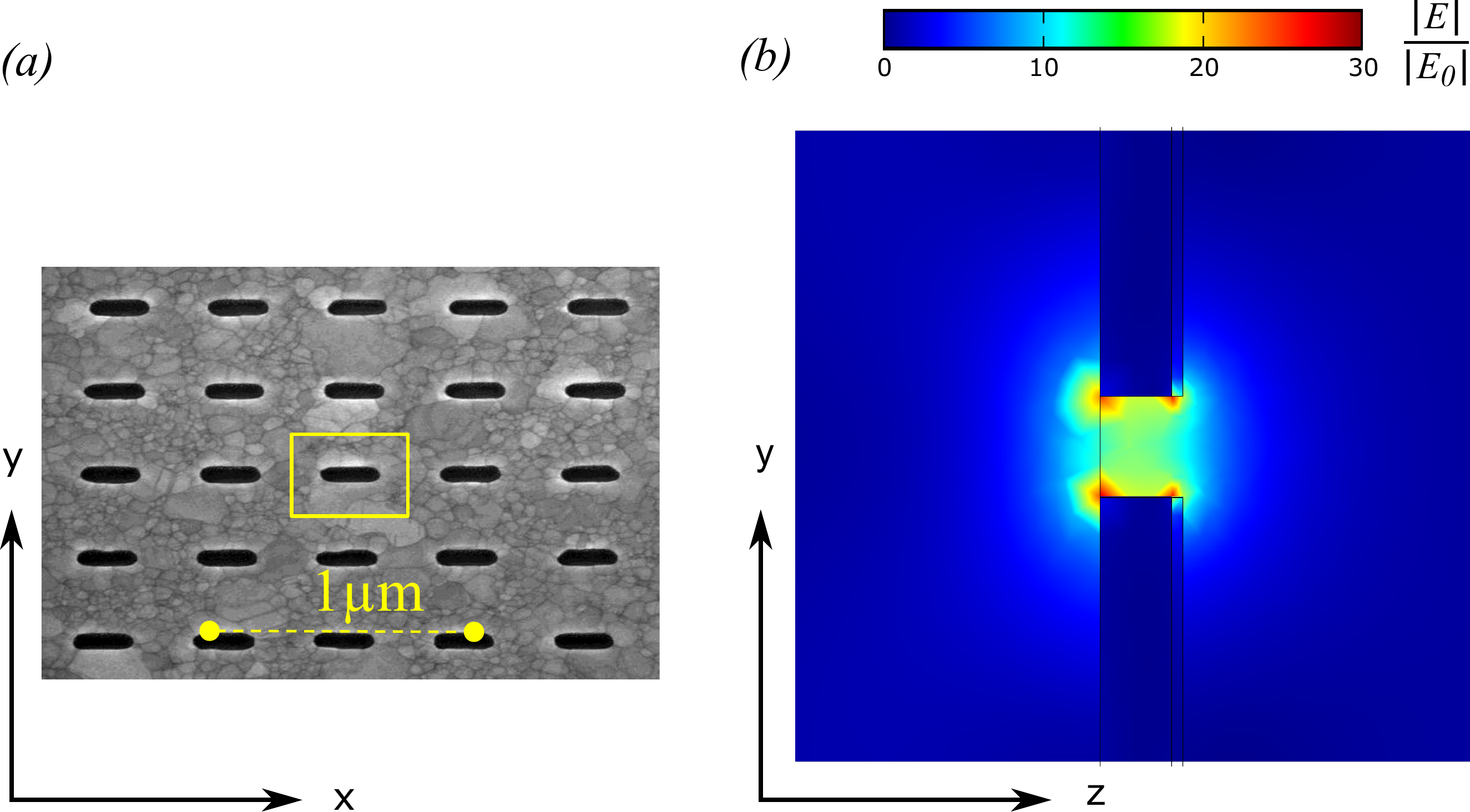}
\caption{(a) SEM image of a fragment of the plasmonic metamaterial with unit cell indicated by the yellow square. (b) Example of the light intensity distribution at the vicinity of one slit (cut view at the center of the slit). The incident beam is coming for the left part of the picture. The caesium vapour is located on the right part of the picture.}
\label{fig2b}
\end{center}
\end{figure}

\begin{figure}
\begin{center}
\includegraphics[width = 0.9\textwidth]{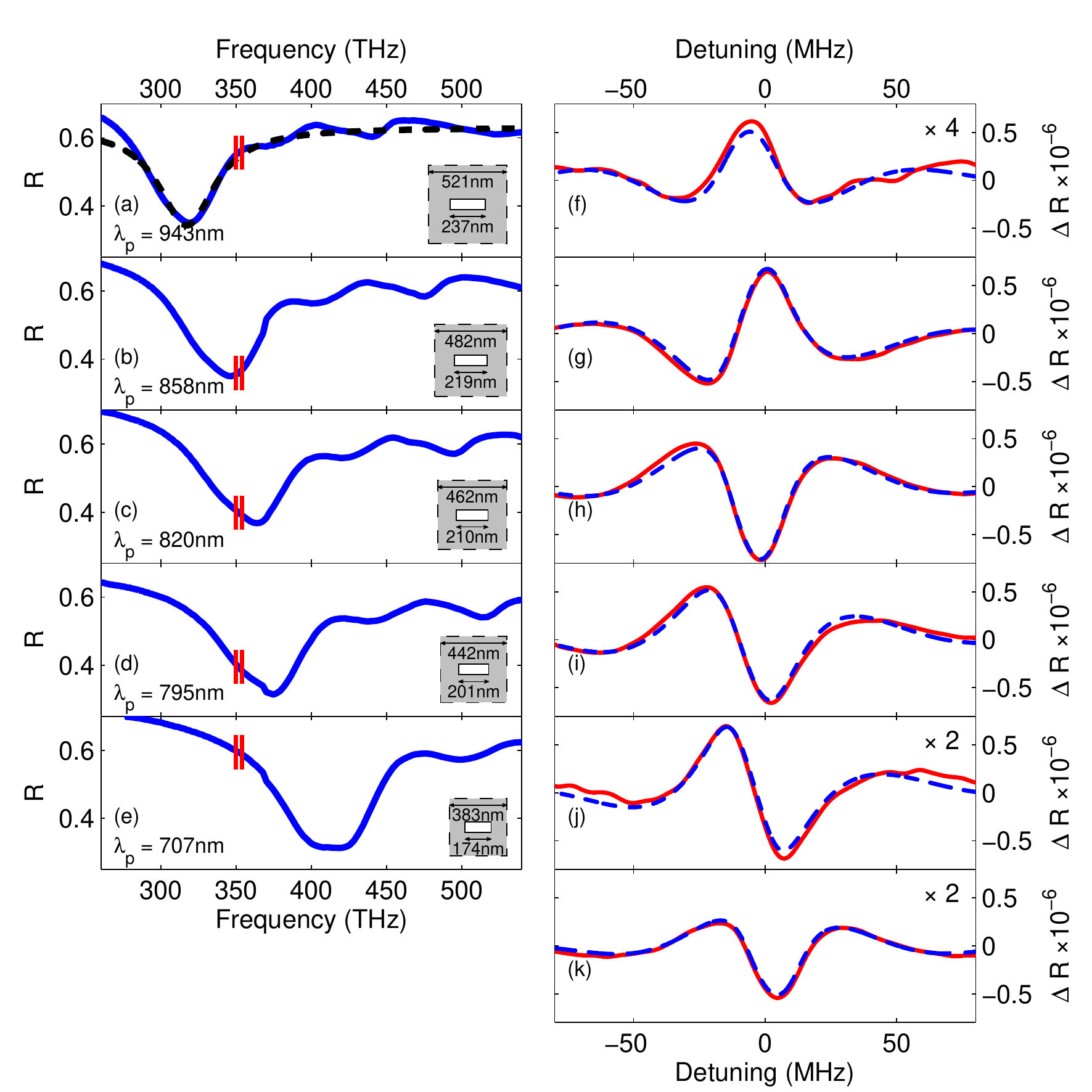}
\caption{Reflection spectra of plasmonic metamaterial/caesium atom vapour hybrid samples. (a)-(e) Broad frequency scan showing the plasmonic resonance lineshape for different metamaterial. The resonance position is indicated on each graph. Insets show sketches of the individual slits of the metamaterial array. The double vertical red lines indicate the position of the D$_2$ line on the plasmon resonance reflectivity dip profile. On (a), the black dashed curve represents a fit of the reflection curve (see supporting information for more details). (f)-(j) Details of the spectra at the caesium D$_2$ line. Each graph is associated to its left inside counterpart. The red curves correspond to the experimental data whereas the blue dashed curves are fits of the relative reflectance $\Delta R$ (see text for more details). The frequency origin corresponds to the resonance of the hyperfine transition $F=4$ to $F=5$ of the caesium D$_2$ line. (k) relative reflectance $\Delta R$ of a dielectric window subtrate, \emph{i.e.} without metamaterial. }
\label{fig3}
\end{center}
\end{figure}

To study modifications of the atomic lineshape in the proximity of metamaterial, we prepare an array of metamaterial samples with different central frequencies of the plasmonic resonance above and below the frequency of the ceasium D$_2$ line. Such metamaterials are fabricated by focused ion beam milling in a silver film of thickness $50$~nm and is post-fabrication protected by $8$~nm layer of SiO$_2$ to prevent chemical attacks to the film. Metamaterial samples, all manufactured on the same window, are two-dimensional arrays of nano-slits $70$~nm wide of varying length from $170$~nm to $240$~nm to achieve plasmonic resonances from $700$~nm to $900$~nm (Fig. \ref{fig2b}). The unit cells are squares with sizes from $w = 380$~nm to $520$~nm long, depending on the length of the slit. Individual reflectance spectra of all metamaterial samples on the array, taken with light polarized perpendicular to the slit, are presented on Fig. \ref{fig3}(a-e). Here dips in the reflectivity spectra coincide with resonant increase of the metamaterial's transmission and absorption.

To study the influence of the nanostructures on the atomic spectra, we use frequency modulation spectroscopy with a single mode diffraction grating-tuneable diode laser with a linewidth of about $100$~kHz. In such measurements, the signal, demodulated in phase with the reference signal at $9$~MHz, is, at a good approximation, proportional to the derivatives of the reflectance spectra. Reflectance spectra are thus obtained after integration of the signal over frequency (see supporting information for more details). Moreover this signal is about seven orders of magnitude more sensitive to the fast variations of reflectivity at narrow atomic vapour lines than to slow variations due to the dispersion of plasmonic resonances. To provide absolute wavelength calibration of the laser reflectometer, we use a second reference caesium cell interrogated in the hole-burning regime where Doppler-free dips provide accurate frequency reference of the hyperfine absorption lines.

Fig. \ref{fig3}(f-j) show the relative reflectance spectra of the caesium $6^2S_{1/2} - 6^2P_{3/2}$ transition between $F = 4$ and $F = 5$ hyperfine sub-states measured for different metamaterial samples. Fig. \ref{fig3}(k) shows the same spectra measured at the plain dielectric window. In all cases, the caesium vapour was maintained at a temperature $T = 80 ^{\circ}$C corresponding to a vapour pressure $P = 7\times 10^{-4}$~mbar and an atoms density of $\rho= 10^{17}$~m$^{-3}$. As only atoms within longitudinal velocity lower than  $\gamma_a/k$ contribute to the sub-Doppler spectra, a $d = k^{-1}$ thick volume of vapour above the unit cell of metamaterial of $w^2$ area contains in average approximately  $\rho w^2 \gamma_a/\bar{v}k^2 \simeq 6 \times 10^{-5}$~atoms contributing to the spectra. Nevertheless, with about $2.5 \times 10^5$ unit cells located and illuminated on each metamaterial sample, the technique allows to record accurate sub-Doppler reflectance spectra of the hybrid atoms/metamaterial device, although less than $20$ vapour atoms contribute to them.

We observe that: 1) the caesium vapour reflectivity spectra, recorded with the presence of metamaterials, are qualitatively different to the spectra recorded at a dielectric window (see Fig. \ref{fig3}(k)); 2) Reflectance spectra of the vapour, recorded at different metamaterial samples, depend on the position of the metamaterials resonance relative to the position of the $6^2S_{1/2} - 6^2P_{3/2}$ line. The latter is shown on Fig. \ref{fig3}(a-e); 3) When the plasmonic resonance is scanned over the atomic line, the reflectivity has a complex lineshape that is characteristic for Fano-type resonances \cite{fano1935sullo,fano1961effects}; 4) Reflectivity curve inverses for positive and negative detuning (compare spectra for plasmonic resonance at $820$~nm and $858$~nm). In these spectra, the spurious contribution of the fluorescence background of the atomic vapour can be ignored. First the fluorescence spectrum is Doppler broadened, thus it appears at a larger frequency span. Second its contribution is weak because of the small solid angle of the detection and because the fluorescence signal is partially removed by the frequency modulation technique.

In our experiment we are dealing with light reflected from the layer of metamaterial followed by an atomic gas layer. Here the plasmonic/atomic system can be modelled as a pair of coupled classical damped harmonic oscillators. The charge displacements $x_p$  and $x_a$  of respectively, the plasmonic and the atomic mode, are governed by the following dynamical equations:

\begin{equation}
\ddot{x}_p+\gamma_p\dot{x}_p+\omega_p^2 x_p+\Omega^2 x_a=-\frac{e}{m}E_0e^{-iwt}.
   \label{eq:xp}
\end{equation}
\begin{equation}
\ddot{x}_a+\gamma_a\dot{x}_a+\omega_a^2 x_a+\Omega^2 x_p=0.
   \label{eq:xa}
\end{equation}
Where $\omega$ is optical laser frequency, $m$ is the effective mass of the charge and $E_0$ represents the incident electric field amplitude; $\gamma_p$ and $\omega_p$ are the damping parameter and the frequency of the plasmonic mode and $\gamma_a$ and $\omega_a$ are the damping parameter and the frequency for the atom. Here $\Omega$ is the coupling constant characterizing the interactions between the light-induced plasmonic and atomic dipoles. We argue that the coupling may be described by a single constant because of spatial averaging over the ensemble of atoms in the near-field layer of the plasmonic metamaterial. We assume that the incident wave drives the plasmonic mode while excitation of the atoms, located at the opposite side of the nanostructure, occurs via the electromagnetic field transmitted through the metamaterial thanks to the resonant excitation of localized plasmonic fields (see Fig. \ref{fig1}).

Equations (\ref{eq:xp}) and (\ref{eq:xa}) of interacting atomic and plasmonic modes describe a typical Fano-like system of a high and low quality coupled modes.  Assuming  $\gamma_p \gg \gamma_a$ we find that (see supplementary materials for a complete derivation) the plasmonic susceptibility remains unaffected by coupling whereas the atomic susceptibility becomes:

\begin{equation}
\chi_a=\rho\frac{6\pi\gamma_a}{k^{3}}\frac{Z}{2(\omega_a-\omega)-i\gamma_a+\Omega_a},
   \label{eq:Xa}
\end{equation}

Compared to $\chi_a^{(0)}=6\pi\gamma_a\rho k^{-3}\left[2(\omega_a-\omega)-i\gamma_a\right]^{-1}$, the atomic susceptibility of the bare atomic sub-Doppler vapour. Here $\Omega_a=\Omega^4\left[\omega(\omega_p^2-\omega^2-i\omega\gamma_p)\right]^{-1}$, where Re$\{\chi_a\}$ and Im$\{\chi_a\}$ correspond to the frequency shift and line broadening induced by the plasmonic coupling respectively while $Z = -\Omega^2\chi_p\omega^2(6\pi\gamma_pNc^3)^{-1}$  describes the plasmon-induced change of the lineshape and mixing of the dispersive and reactive parts of the atomic response. $N$ is the free electrons density participating to the plasmonic resonance.

We analyze the relative reflectance spectra by doing a least square non linear fit using the atomic susceptibilities of Equation \ref{eq:Xa}, with $\gamma_a$, $Z$ (real and imaginary parts), and $\Omega_a$ as the fitting parameters. The fits correspond to the blue dashed line in Fig. \ref{fig3}(f-k). They are in very good agreement with the experimental data. This allows the evaluation of the coupling constant $\Omega$ and the atomic linewidth  $\gamma_a$ for all plasmonic nanostructures. The corresponding coupling frequency was found to be $\Omega = 0.01\sqrt{w_a\gamma_p} \simeq 1$~THz. It is much higher than the atomic natural linewidth $\gamma_a/2\pi=5.2$~MHz, leading to a strong modification of the line shape. However, its effect on the atomic line frequency shift that is given by $|\Omega_a |/2\pi\simeq 0.3$~MHz, is small in comparison to the bare transition linewidth $\gamma_a/2\pi$, to be reliably detected, as confirmed by our experiments.

For $|\Omega_a|\ll\gamma_a$ dispersion of the atomic response (and thus the atomic contribution to the reflectance spectra) is well approximated by a product of undisturbed atomic and plasmonic responses, as can be seen from the expression for the atomic susceptibility, as follows from the Equation (\ref{eq:Xa}):

\begin{equation}
\chi_a=-\rho\frac{\gamma_a \Omega^2}{\gamma_pN\omega}\frac{\chi_p}{2(\omega_a-\omega)-i\gamma_a}.
   \label{eq:Xa2}
\end{equation}

Since the frequency variations of the atomic spectra are much faster than that of plasmonic spectra, the resulting atomic response is proportional to $\chi_p$, which is a complex multiplier constant. This complex multiplier mixes up the real and imaginary parts of the atomic response radically changing the reflectivity spectra without a noticeable shift of the atomic line.

Finally, we found that the atomic linewidth in the presence of metamaterial nanostructure is $\gamma_a \simeq 20$~MHz. This value is much smaller than the Doppler frequency broadening of 200 MHz but it is larger than the natural atomic linewidth of $5.2$~MHz. We note that similar linewidth was also seen on the window subtrate (See Fig. \ref{fig3}) and therefore the Purcell mechanism of line broadening related to the increased density of state in the localized plasmonic mode can be ruled out, arguably due to weak effective mode confinement. Thus we conclude that the broadening is likely due to the presence of a small residual velocity component along the propagation axis of incidents light of the atoms traveling at the vicinity of the nanostructure.

In conclusion we report new plasmo-atomic metamaterial with Fano sub-Doppler spectra. We provide experimental evidence that hyperfine sub-Doppler spectra of atomic gas are modified in the presence of a nanostructured plasmonic metamaterial which has resonances millions of times broader than atomic lines. We show that strong modification of atomic spectra in the nearfield of metamaterial depends on mutual position of atomic and plasmonic resonances. Our results imply that such highly-sensitive atomic probe can be used for detailed characterization of the evanescent field and the plasmonic dispersion of the nanostructures. For example, one can use an atomic quadrupole transition which is sensitive to the local electric field gradient\cite{deguchi2009simulation,kern2012strong}. Moreover, the ability to tailor the atomic lineshapes may have implications for the nanotechnology-enabled atom-based metrology, sensing, and the development of atom-on-a-chip applications. One way to achieve those tasks would be to laser cool the atoms before transferring them at the vicinity of the metamaterial. Then the field pattern, generated by the metamaterial, could be used as a platform to trap the cold gas \cite{gullans2012lattices}. This configuration allows to have greater control on the atoms/metamaterial coupling and should greatly increase the signal-to-noise ratio necessary for metrology and sensing applications. Furthermore, we shall expect that luminescence and nonlinearity of the atomic vapour to be enhanced in the hybrid systems while plasmo-atomic interaction can lead to long-range lattice effects on the trapped atoms for quantum simulations \cite{gonzalez2015lattices}.

\begin{suppinfo}
Details about the experimental apparatus, samples,  the derivation of the reflectance signal and derivation of the susceptibilities are presented. The authors declare no competing financial interest.\end{suppinfo}

\begin{acknowledgement}
All authors acknowledge the support of the Singapore Ministry of Education Academic Research Fund Tier 3 (Grant No. MOE2011-T3-1-005) for funding equipments and consumable. N.I.Z. thanks EPSRC (UK) via the Programme on Nanostructured Photonic Metamaterials EP/G060363/1. M.D. thanks the Physics and Applied Physics division of the School of Physical and Mathematical Science at NTU for its support during his stay in Singapore.
\end{acknowledgement}

\section{Contributions}
The idea of experiment belongs to M.D., N.I.Z. and D.W. The spectroscopic approach was devised by M.D. The experimental apparatus was assembled by S.A., A.E.C. and G.A. who also conducted the measurements and developed and fabricated metamaterial samples. S.A., A.E.C. and G.A. provided equal contribution to this work.  D.W. developed theoretical model. N.I.Z and D.W. wrote the manuscript. All authors contributed to the interpretation of results and revised the manuscript. N.I.Z. supervised the work.


\providecommand{\latin}[1]{#1}
\providecommand*\mcitethebibliography{\thebibliography}
\csname @ifundefined\endcsname{endmcitethebibliography}
  {\let\endmcitethebibliography\endthebibliography}{}




\section*{Supplementary material}

# Supporting information for:
# Atomic Response in the Near-field of Nanostructured Plasmonic Metamaterial

Syed Abdullah Aljunid,[†] Eng Aik Chan,[†] Giorgio Adamo,[†] Martial Ducloy,[*,†,‡,¶]
David Wilkowski,[*,†,¶,§,∥] and Nikolay I. Zheludev[*,†,⊥]

[†]*Centre for Disruptive Photonic Technologies, TPI, Nanyang Technological University, 637371, Singapore.*

[‡]*Laboratoire de Physique des Lasers, Université Paris 13, CNRS, (UMR 7538), F-93430, Villetaneuse, France.*

[¶]*School of Physical and Mathematical Sciences, Nanyang Technological University, 637371 Singapore.*

[§]*Centre for Quantum Technologies, National University of Singapore, 117543 Singapore.*

[∥]*MajuLab, CNRS-UNS-NUS-NTU International Joint Research Unit UMI 3654, Singapore.*

[⊥]*Optoelectronics Research Centre & Centre for Photonic Metamaterials, University of Southampton, Southampton SO17*

E-mail: martial.ducloy@univ-paris13.fr; david.wilkowski@ntu.edu.sg; nzheludev@ntu.edu.sg

## Experimental apparatus and samples

Caesium vapour is kept in a vacuum chamber with metamaterials sample fabricated on one of the optical windows of the chamber. A quartz wedge window (3°) is used to support a 50 nm film silver deposited by thermal evaporation and protected by 8 nm layer of $SiO_2$.

Several metamaterials samples were fabricated on the film by focused ion beam milling. Each metamaterials array is 200μm × 200μm in size. We also fabricate a window in the silver film for reference measurements. Before releasing the caesium in the chamber, it is pumped to $10^{-8}$ mbar. The caesium vapour is maintained at temperature of 80 °C corresponding to vapour pressure $P = 7 \times 10^{-4}$ mbar and vapour density of $\rho = 10^{17}$ m$^{-3}$.

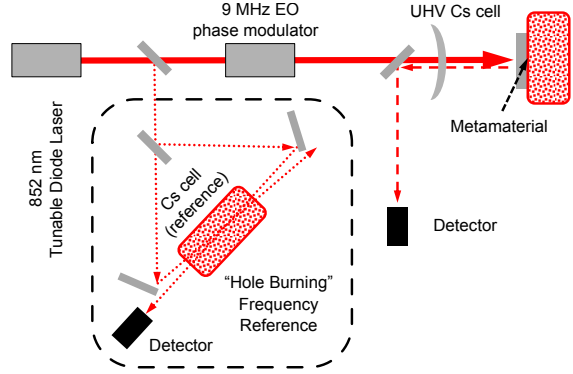

Figure S1: Schematic representation of the optical setup, including frequency reference attachment.

We use an 852 nm grating-tuneable semiconductor diode laser with line width lower than 100 kHz (TOPTICA model DLPro100) as the optical source in the experiment (see Fig. S1). The laser has a mode hope-free spectral tuning of 13 GHz which allows scanning of the complete hyperfine structure of the dipole allowed D$_2$ line as well as the Doppler spectrum. An intensity of $I = 2$ mW/cm$^2$ is used for the reflectance measurements of the hybrid metamaterial. This value is below the saturation intensity ($I_s = 2.7$ mW/cm$^2$). We note that no saturation effect has been observed. The unused light intensity is redirected to a spectroscopic reference setup. For reflectance measurements the laser beam is phase-modulated at 9 MHz with an electro-optical modulator before being focused at normal incidence by a 20 cm lens at the surface of the window containing metamaterial films. The reflected signal is extracted with a beam-splitter (10:90) and detected by an avalanche silicon photodetector (HAMAMATSU model C10508). The reflection signal is demodulated and amplified using a lock-in amplifier (Zurich Instruments model HF2LI) while scanning, at

low frequency, the optical frequency of the laser across the absorption line. In addition through-transmission of the vapour sample is also continuously measured to monitor the pressure of caesium vapour. To provide absolute optical frequency calibration of the laser reflectometer, we use an extra caesium cell, at room temperature, interrogated in saturated absorption spectroscopy where Doppler-free dips in the probe transmission provides reference to the caesium hyperfine absorption lines. The accuracy of this measurement is better than 100 kHz.

## Derivation of the reflectance signal

In a multi-layered system comprising of window substrate, plasmonic metamaterial and atomic layer with respective refractive indexes $n$, $n_p$ and $n_a$, the amplitude reflectance is given by

$$r = \frac{r_1 + r_2 e^{2in_p l\omega/c}}{1 + r_1 r_2 e^{2in_p l\omega/c}}, \qquad (1)$$

where $r_1 = (n - n_p)/(n + n_p)$ and $r_2 = (n_p - n_a)/(n_p + n_a)$. Since the atomic medium is diluted, the atomic susceptibility $\chi_a \ll 1$, thus allowing for the expansion $r \simeq r|_{\chi_a=0} + \partial r/\partial \chi_a |_{\chi_a=0} \chi_a$, where $n_a \simeq 1 + \chi_a/2$. The reflected intensity coefficient can then be presented as:

$$R(\omega) = R|_{\chi_a=0} + \Delta R(\omega), \qquad (2)$$

Where

$$R|_{\chi_a=0} = |r|^2_{\chi_a=0} \qquad (3)$$

and

$$\Delta R(\omega) = 2\text{Re}\left\{\left(r^* \frac{\partial r}{\partial \chi_a}\right)|_{\chi_a=0} \chi_a(\omega)\right\} \qquad (4)$$

Here $R|_{\chi_a=0} \equiv R$ is measured independently (see Fig. 3(a-e) in the letter).

To study the influence of the nanostructures on the atomic spectra we used frequency modulation spectroscopy technics. In such measurements, the intensity of the component of

the reflectivity signal that is in-phase with the modulation signal at $\omega_m$, takes the following form:

$$I_{phase} \simeq \text{Re}\left\{\left(r^*\frac{\partial r}{\partial \chi_a}\right)|_{\chi_a=0}[\chi_a(\omega+\omega_m)-\chi_a(\omega-\omega_m)]\right\}, \quad (5)$$

Assuming $\chi_a(\omega+\omega_m)-\chi_a(\omega-\omega_m) \simeq 2\partial\chi_a/\partial\omega_m$, the reflectance spectrum is extracted as:

$$\Delta R(\omega) = \frac{1}{\omega_m}\int I_{phase}\mathrm{d}\omega \quad (6)$$

The problem reduces now to the determination of the susceptibilities of the atomic vapour $\chi_a$ and of the metamaterial entering in the expression of the reflectance $r$.

## Two-coupled oscillator model: derivation of the susceptibilities

In the case of the dielectric window substrate with no metamaterial, the calculation is straightforward: The Doppler-free atomic susceptibility reduces to $\chi_a^{(0)} = 6\pi\gamma_a\rho c^3 k^{-3}\left[2(\omega_a-\omega)-i\gamma_a\right]^{-1}$ and the amplitude reflectance is simply $r = (n-n_a)/(n+n_a)$, where $\rho$ is the atomic density, $k$ is the wavevector and $\gamma_a$ is the linewidth of the transition. We note that the expression of the atomic susceptibility corresponds to a bulk system, where surface interaction are not considered[S1]. For the complete plasmono-atomic system, we use a two-coupled classical harmonic model to derive the plasmonic and atomic susceptibilities. Our starting point is the set of dynamical equations of the charge displacement for the plasmonic and the atomic mode [see Eqs. (1) and (2) in the letter]. At the vicinity of the atomic resonance, i.e. $\omega \simeq \omega_a$, assuming $\gamma_p \gg \gamma_a$, $\Omega^4\omega_a^{-2}\gamma_a^{-1}$, the broad plasmonic resonance remains unaffected by the coupling, but not the atomic resonance. Indeed, solutions of Eq. (1) and (2) in the letter are:

$$x_p = -\frac{eE_0}{m}\frac{1}{\omega_p^2-\omega^2-i\omega\gamma_p}e^{i\omega t} \quad (7)$$

$$x_a = x_p\frac{\Omega}{\omega}\frac{1}{2(\omega_a-\omega)-i\gamma_a+\Omega_a} \quad (8)$$

Here $\Omega_a = \Omega^4 \left[\omega(\omega_p^2 - \omega^2 - i\omega\gamma_p)\right]^{-1}$, where $\text{Re}\{\chi_a\}$ and $\text{Im}\{\chi_a\}$ correspond to the frequency shift and line broadening induced by the plasmonic coupling. From Eq. (7), the metamaterial index of refraction can be can be written as $n_p\sqrt{1 + \chi_b + \chi_p}$, where

$$\chi_p = \frac{Ne^2}{m} \frac{1}{\omega_p^2 - \omega^2 - i\omega\gamma_p} \tag{9}$$

and $\chi_b = 29 + i$ is the background susceptibility of the bulk silver at 852 nm and $N$ is the free electrons density participating to the plasmonic resonance. From here we compute intensity reflectance spectra $R$ of the metamaterial and fit them to the experimental measured spectra to extract the values of $\gamma_p$ and $\omega_p$. An illustrative example of such a fit is presented by the black dashed line in Fig. 3(a) of the letter. We find $\gamma_p \simeq 35$ THz. The values of $\lambda_p = 2\pi c/\omega_p$ are given for each plasmonic structure in Fig. 3(a-e) of the letter. Using the extracted values of $\gamma_p$, and $\omega_p$ the susceptibility of the atomic vapour is calculated as:

$$\chi_a = \rho \frac{6\pi\gamma_a}{k^3} \frac{Z}{2(\omega_a - \omega) - i\gamma_a + \Omega_a} \tag{10}$$

where $Z = -\Omega^2\chi_p\omega^2(6\pi\gamma_p Nc^3)^{-1}$ mixes the dispersive and reactive parts of the atomic response. From here the hybrid reflectivity spectra can be fitted to the experimentally measured dispersions, showing very good agreement with the experiment [Fig. 3(f-h) in the letter]. This allows the evaluation of the coupling frequency $\Omega$ and the atomic linewidth $\gamma_a$ for all plasmonic nanostructures.

# References

(S1) L. Allen and J. H. Eberly, *Optical resonance and two-level atoms* (Courier Corporation, 2012).



\end{document}